\documentclass[aps,pre,twocolumn,groupedaddress,floatfix]{revtex4}   
\usepackage[utf8]{inputenc}
\usepackage{amsmath,graphicx}
\usepackage[usenames]{color}

\newcommand{\be}{\begin{equation}}
\newcommand{\ee}{\end{equation}}
\newcommand{\ba}{\begin{eqnarray}}
\newcommand{\ea}{\end{eqnarray}}

\begin{document}

\title{Structure, diffusion and orientational freezing in lithium metasilicate}

\author{Cristian Balbuena}
\affiliation{Universidad Nacional del Sur, Departamento de Química — INQUISUR, Av. Alem 1253, 8000-Bahía Blanca, Argentina}

\author{Carolina Brito}
\affiliation{Departamento de F\'{\i}sica,
Universidade Federal do Rio Grande do Sul,
CP 15051, 91501-970 Porto Alegre, RS, Brazil}

\author{Daniel A. Stariolo}
\email{daniel.stariolo@ufrgs.br}
\affiliation{Departamento de F\'{\i}sica,
Universidade Federal do Rio Grande do Sul,
CP 15051, 91501-970 Porto Alegre, RS, Brazil}

\begin{abstract}
We report on the dynamic and structural characterization of lithium metasilicate $Li_2SiO_3$, a network forming ionic glass,
by means of molecular dynamics  simulations. The system is characterized by  a network of $SiO_4$ tetrahedra disrupted by $Li$
ions which diffuse through the network. Measures of mean square displacement of $Si$ and $O$ atoms allow us to identify a
temperature at which tetrahedra stop moving relative to each other. This temperature $T_c\approx 1500\,K$ can be characterized within the
framework of mode coupling theory. At a much lower temperature $T_g\approx 1000\,K$, a change in the slope of the volume versus
temperature data allows to single out the glass transition. We find signatures of both transitions in structural order parameters,
related to the orientation of tetrahedra. Going down in temperature we find that, around the mode coupling transition temperature,
a set of order parameters which measure the relative orientation of tetrahedra cease to increase and
stay constant below $T_c$. Another well known measure of orientational order, the bond orientational order parameter, which in the
studied system measures local order
within single tetrahedrons, is found to continue growing below $T_c$ until $T_g$, below which it remains constant. Our results allow
to relate two characteristic dynamic transitions with corresponding structural transitions, as observed in two different orientational
order parameters. Furthermore, the results indicate that the network of thetrahedra continue to relax well below the point where neighboring tetrahedra cannot rearrange relative to each
other, and the glass is reached only upon a process of relaxation of atoms which form the thetrahedron, as quantified by the change in the bond orientational order parameters.
\end{abstract}
\maketitle

\section{Introduction}

Silicate oxide glasses are the most used and studied kind of inorganic glasses. 
They have many applications ranging from window glasses to optical fibres \cite{Greaves2007}.
Typically, the local structure of these systems are tetrahedra of silicons ($Si$) and  oxigens ($O$),  which are connected among them
forming a network. The presence of lithium atoms ($Li$), which are called  modifying cations, 
disrupts the connectivity of some tetrahedra
and are responsible for important physical properties of the system, as for example electric conduction \cite{Dyre2009, Lammert2010, Greaves2007}.
$SiO_4$ tetrahedra are formed at already high enough temperature, where they diffuse toghether with the free $Li$ atoms.
This is the liquid state of the system of lithium metasilicate $Li_2SiO_3$. 
As temperature decreases the atoms of $Si$ an $O$ stop diffusing but the $Li$ keeps traveling in the network of tetrahedra.
This sharp slowdown in the dynamics of the tetrahedra allows to define the glass transition temperature $T_g$. 
Below $T_g$ the system displays an ionic glass phase.

In the last decades much effort has been made to understand the nature of the glass transition and characterize the glassy phase in many different systems \cite{HoKoBi2002, SaAn2003, BerthierRMP2011, Greaves2007}. One important question is if there is any structural signature accompanying the strong
slowing down of the dynamics as the system appraches the glass transition. While there is no definitive answer to this question yet, the search
for such structural signatures (besides the obvious relevance to the whole picture of the glass transition) is important as it gives new insights
into the nature of the slowing down. Among the dynamical approaches, one of the most accepted theories to describe the transition  is the Mode-Coupling theory (MCT). Although this is an approximate theory,  MCT  accounts well for some aspects of the initial slowing down of the dynamics, in particular the appearance of a two step relaxation of correlations with a non trivial short time $\beta$ regime and a structural, long time $\alpha$ regime
 \cite{Gotze1999,Berthier2010, BiroliBouchaud2012} . It predicts a critical-like divergence of relaxation times at a characteristic temperture $T_c$.
One important limitation of MCT is the fact that it cannot capture important features of a system when the liquid state  has short-range 
structural order \cite{Tanaka2012}.  This is because the theory is based on the behavior of the density correlation funtion of the system. This is enough to describe an homogeneous liquid. However, systems like lithium metasilicate have an important local structure given by a network of tetrahedra.
The short-range order  of tetrahedra is very robust even in the liquid phase, posing strong orientational constraints to relaxation. In this case, it seems 
important to use a description that is able to capture this short-range order and analize its influence in the dynamical properties of the system.

In this work we report a detailed study of dynamical and structural properties of $Li_2SiO_3$ using two complementary approaches: (i) we first analyze the dynamics by mesuring the mean square displacement of each kind of atom and the self-correlation function of its positions. We analyze these quantities in the context of MCT and extract the diffusion coefficient and relaxation time, which allow us to define the critical mode-coupling temperature $T_{c}$.  (ii) Then we measure and analyze two orientational quantities. The bond-orientational or Steinhardt order parameters allow to identify the local structure of the network and its robustness 
to temperature changes \cite{Steinhardt1983}. To go beyond this local measure and  analyze the relative orientations of the local structure in space, we measure a recently defined quantity, called ``Rey parameters'' \cite{Rey2007}.
We show that both descriptions, the dynamic and the structural, are quite complementary. The  two characteristic temperatures extracted from dynamics, namely $T_{c}$ and $T_g$, have an interpretation in terms of the orientaional parameters:
$T_{c}$ correspons to the temperature below which the relative orientation {\it between} tetrahedra (Rey parameters) ceases,  and $T_g$ corresponds to the temperature below which the average positions of the atoms that compose the tetrahedra do not change any more.
This allows to go beyond MCT and give a spatial interpretation of the events that happen between  $T_{c}$ and $T_g$.

The paper is organized as follows: in section \ref{sec.model} we define the model and details of the molecular dynamics protocol. In section
\ref{sec.dynamics} we analyze the dynamics of the system in the context of mode coupling theory. In section \ref{sec.orientational} we define
and analyze orientational order parameters. In sec. \ref{sec.conclusions} we end with a discussion of our results.
\section{The model} \label{sec.model}
We have performed classical molecular dynamics simulations of $Li_2SiO_3$ containing $N=3456$ particles, of which $N_{Li}=1152$, $N_{Si}=576$, $N_O=1728$ correspond to lithium, silicon and oxigen atoms respectively.
The interaction potential was chosen to be of the Gilbert-Ida type~\cite{Ida1976} with and $r^6$ term:
\be \label{potential}
U_{ij}=\frac{e^2}{4\pi \epsilon_0}\frac{q_iq_j}{r}-\frac{c_ic_j}{r^6}+
                     f_0(b_i+b_j)\exp{\left[\frac{a_i+a_j-r}{b_i+b_j}\right]}
\ee
where the pair of indexes $i,j$ characterizes different pairs $Li$-$Li$, $Li$-$Si$, $Li$-$O$, etc. and $r$ is the corresponding distance. The first term in (\ref{potential}) is the Coulomb interaction with the effective charges
$q_i$ for $Li$, $Si$ and $O$. The long-range nature of the Coulomb interaction was taken
into account by means of Ewald summation method. The second term is a dispersive interaction and the last term is a Born-Meyer type potential that  accounts for repulsive short-range interactions. The potential parameters were attributed on the basis of ab-initio
molecular orbital calculations~\cite{HaOk1992}. Their validity was checked in the liquid, glassy and crystal states under constant pressure conditions showing good agreement with experimental data~\cite{HeKuVoBa2002,BaHe2001}. The simulations were implemented with the
software LAMMPS~\cite{Pl1995} and a Verlet algorithm with a time step of $1\, fs$ was used to integrate the equations of motion. The box size was determined by performing simulations in the NPT ensemble at atmospheric pressure, which allows us to reach the same density as the experimental data for any temperature within the experimental error.

The system was equilibrated at $3500\, K$ for more than $0.5\, ns$, starting from a random configuration, using the NVE ensemble. After that, the temperature was decreased until $3200\, K$ and followed by $2\, ns$ in NPT, $1\, ns$ in NVE and $1\, ns$ in NVT ensemble to ensure the absence of a drift in pressure and temperature. Then, the system was equilibrated in a $2\, ns$ run using the NVE ensemble. Finally, after this equilibration procedure, trajectories of $1\, ns$ were generated in the NVE ensemble for analysis. 

In this work we present simulations for a wide range of temperatures $600\,K \leq T \leq 3200\, K$. At each temperature the system was prepared following the same protocol before starting the production run.

\begin{figure}
\includegraphics[width = 1.0\columnwidth]{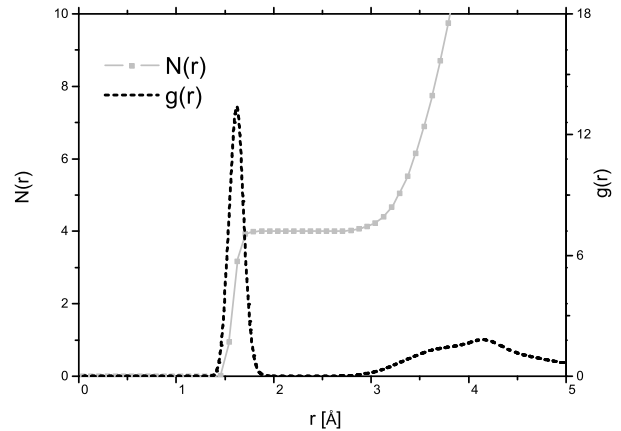}
\caption{The radial distribution funcion $g(r)$ for Si-O pairs in lithium metasilicate and the coordination number $N(r)$.}
\label{gr_Si0}
\end{figure}

Structurally, ionic silicates are formed by a network of $SiO_4$ tetrahedra in which silicon is surrounded by four oxygen atoms, disrupted by the presence of modifier alkali atoms.
Then, the glassy matrix is a result of the chain formation of these tetrahedra sharing their apices (bridging oxygens $BO$). The incorporation of a small amount of $Li_2O$ (modifier oxide) does not destroy the $SiO_4$ tetrahedra but replaces a covalent bridge ($BO$) by  non-bridging oxygen atoms ($NBO$) \cite{Balbuena2013}. 
In Fig. \ref{gr_Si0} we show the radial distribuition
function (RDF) $g(r)$ between atoms of $Si$ and $O$. The integral of  $g(r)$ is the coordination number of the $Si$ atoms:
\be
N(r)=4\pi \rho \int_0^r r^2\,g(r)\,dr,
\ee
where the existence of 4 neighboring oxygens is evident. 

\section{Dynamic characterization of lithium metasilicate} \label{sec.dynamics}
\begin{figure}
\includegraphics[width = 1.0\columnwidth]{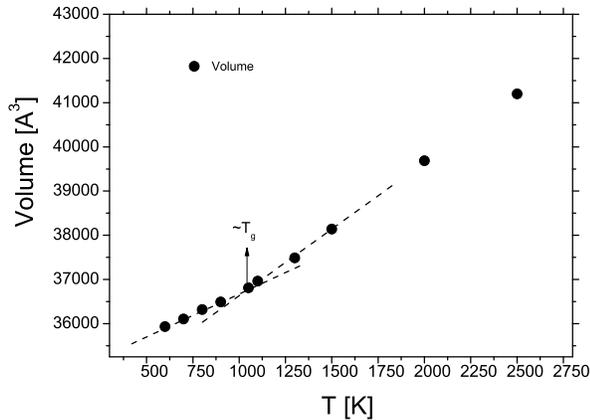}
\caption{Free volume as a function of temperature. A change in the the slope of the curve occurs at $T_g \approx 1000K$ signaling a glass transition.}
\label{glassT}
\end{figure}

Our simulations of lithium metasilicate show the presence of a glass transition at $T_g \approx 1000K$,  as can be seen in  Fig. \ref{glassT} by the change 
of the specific volume as funcion of temperature. Similar results were obtained by other authors \cite{habasakiTg2004,BaHe2001} . 

\begin{figure*}[h,t]
\centering
\includegraphics[width = 0.6\columnwidth, height=0.5\columnwidth]{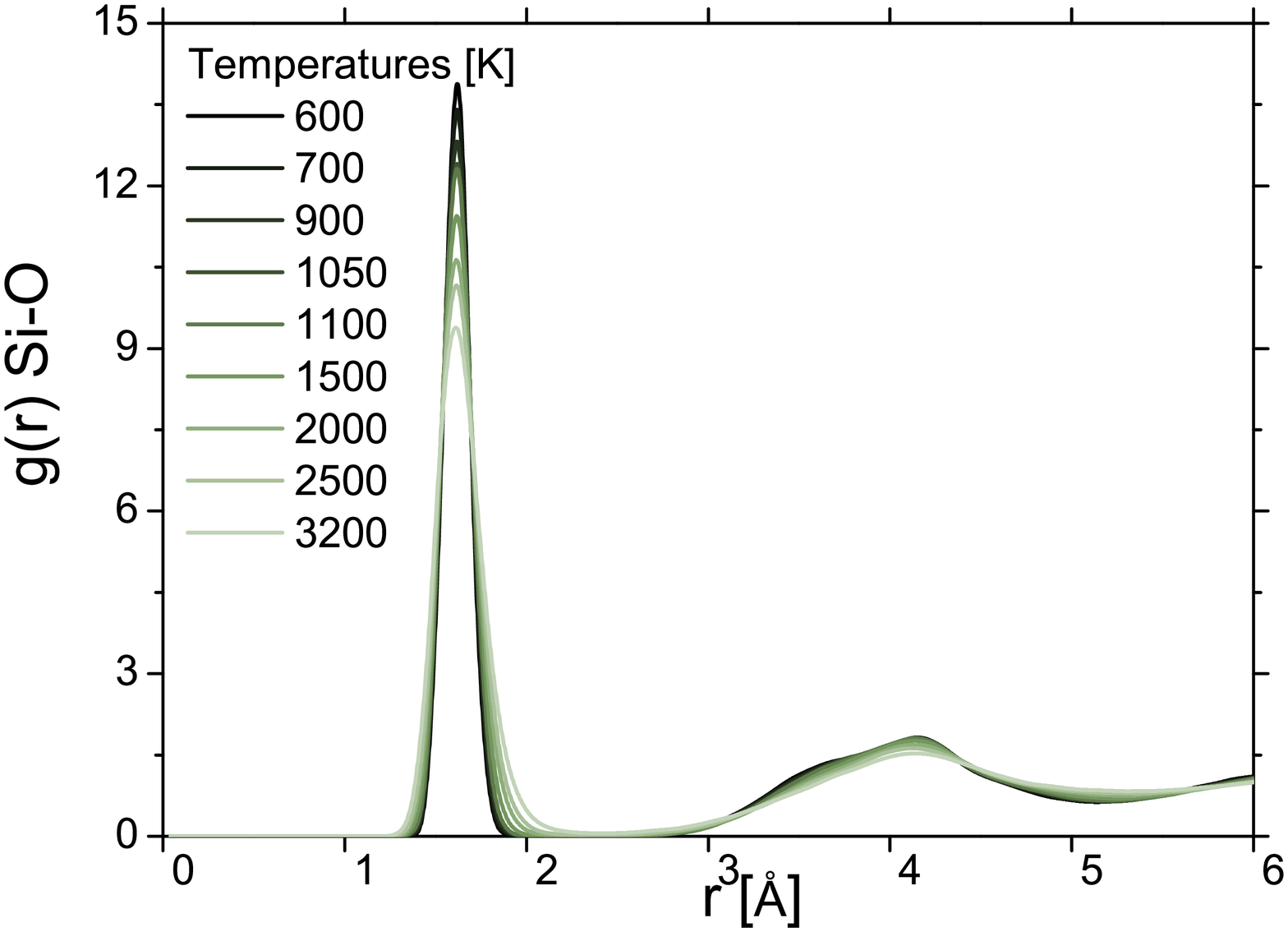}
\includegraphics[width = 0.6\columnwidth, height=0.5\columnwidth]{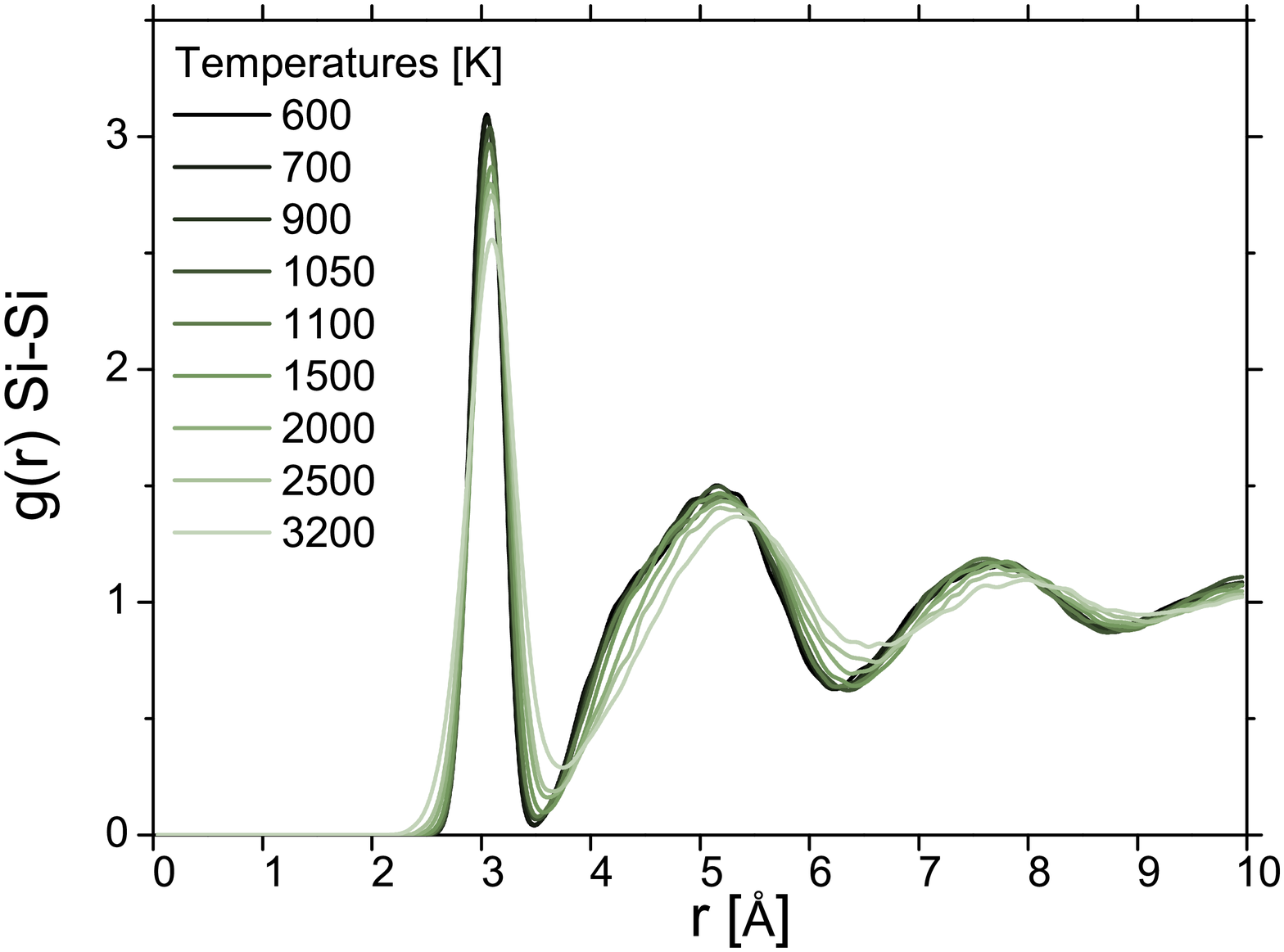}
\includegraphics[width = 0.6\columnwidth, height=0.5\columnwidth]{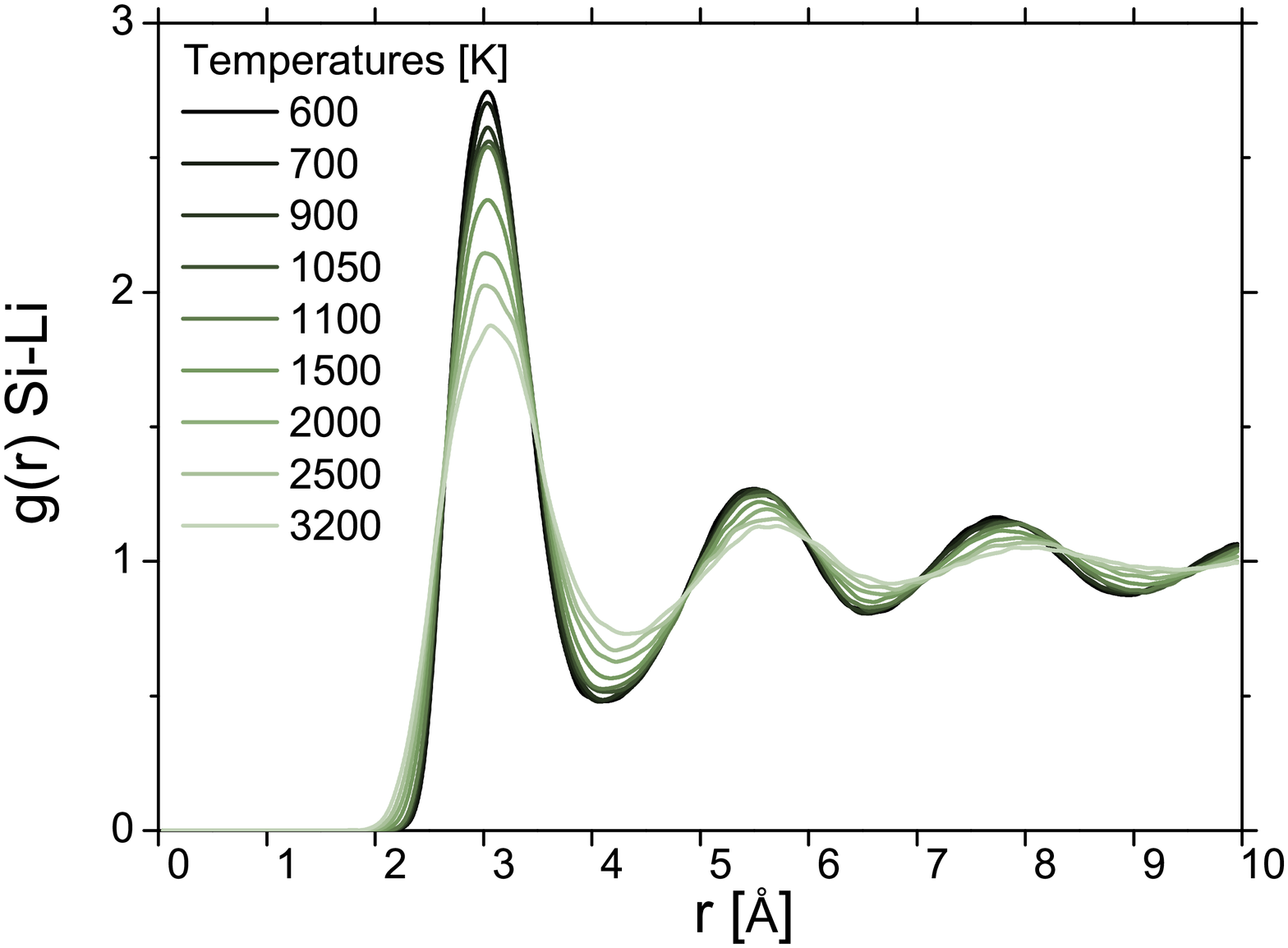}
\caption{Radial distribution function between different species of atoms for the range of temperatures $600\,K\leq T\leq 3500\,K$. Left: Si-O,
 center: Si-Si, right: Si-Li.}
\label{gr}
\end{figure*}

\begin{figure*}[h,t]
\centering
\includegraphics[width = 0.6\columnwidth, height=0.5\columnwidth]{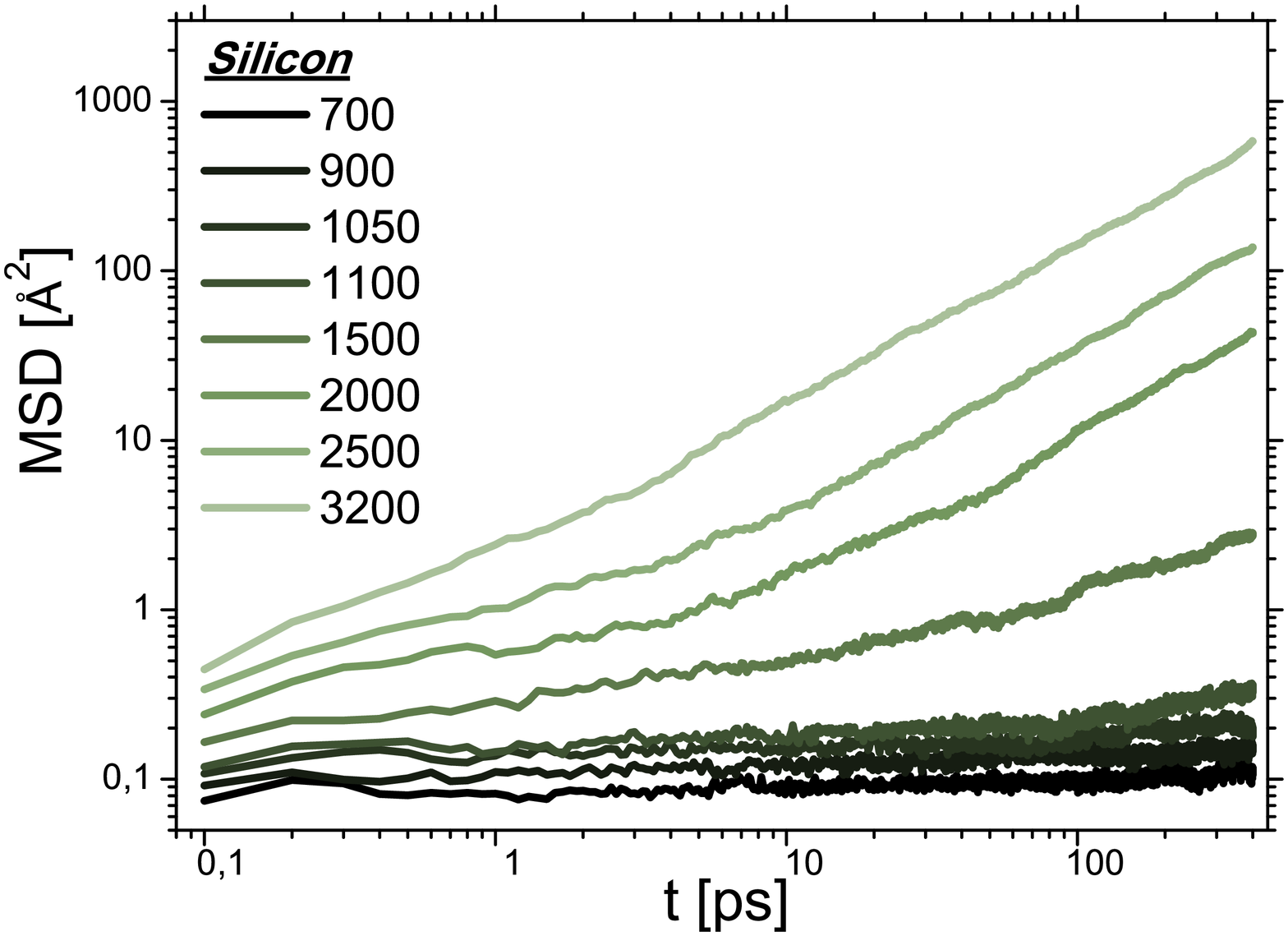}
\includegraphics[width = 0.6\columnwidth, height=0.5\columnwidth]{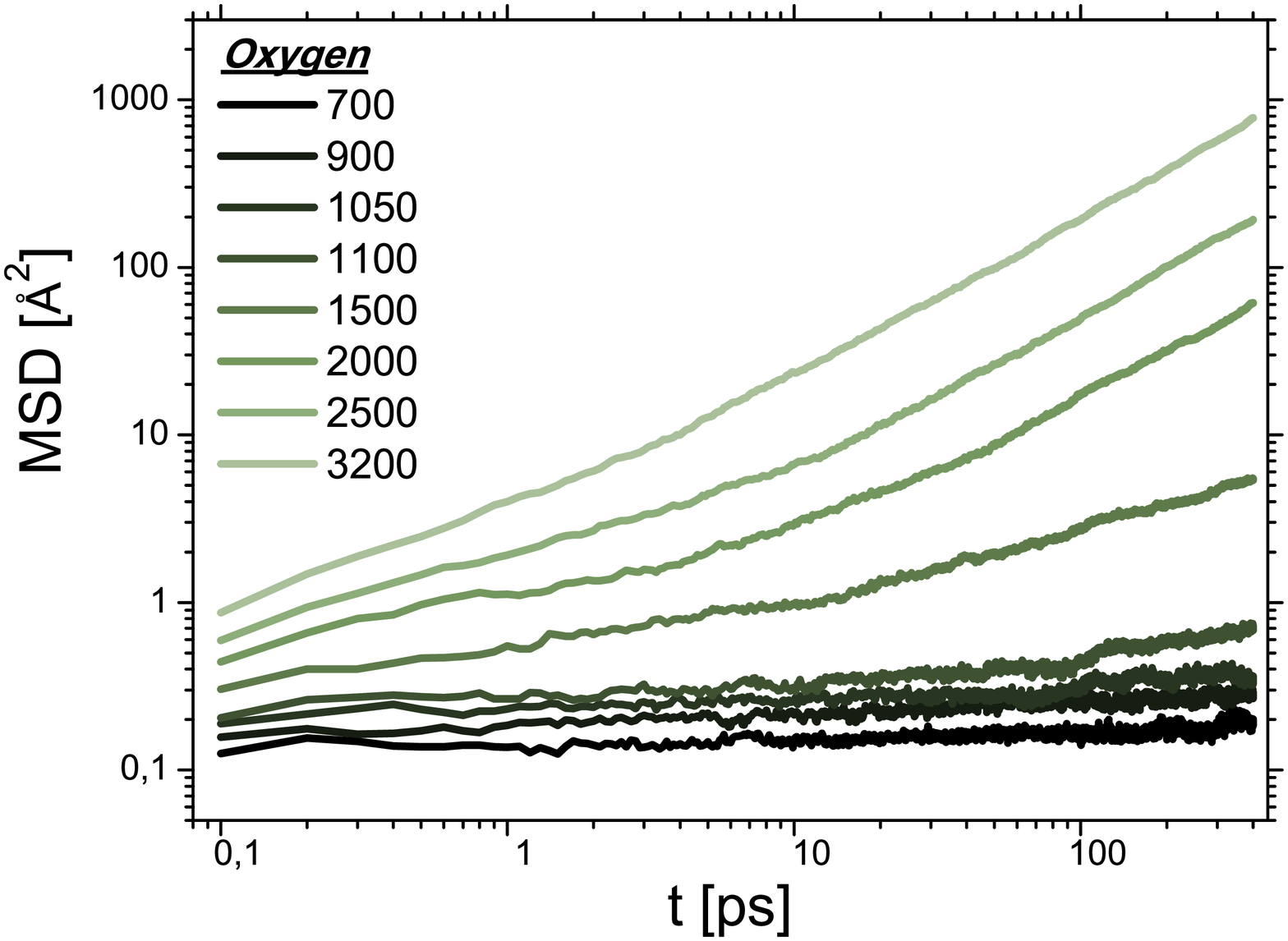}
\includegraphics[width = 0.6\columnwidth, height=0.5\columnwidth]{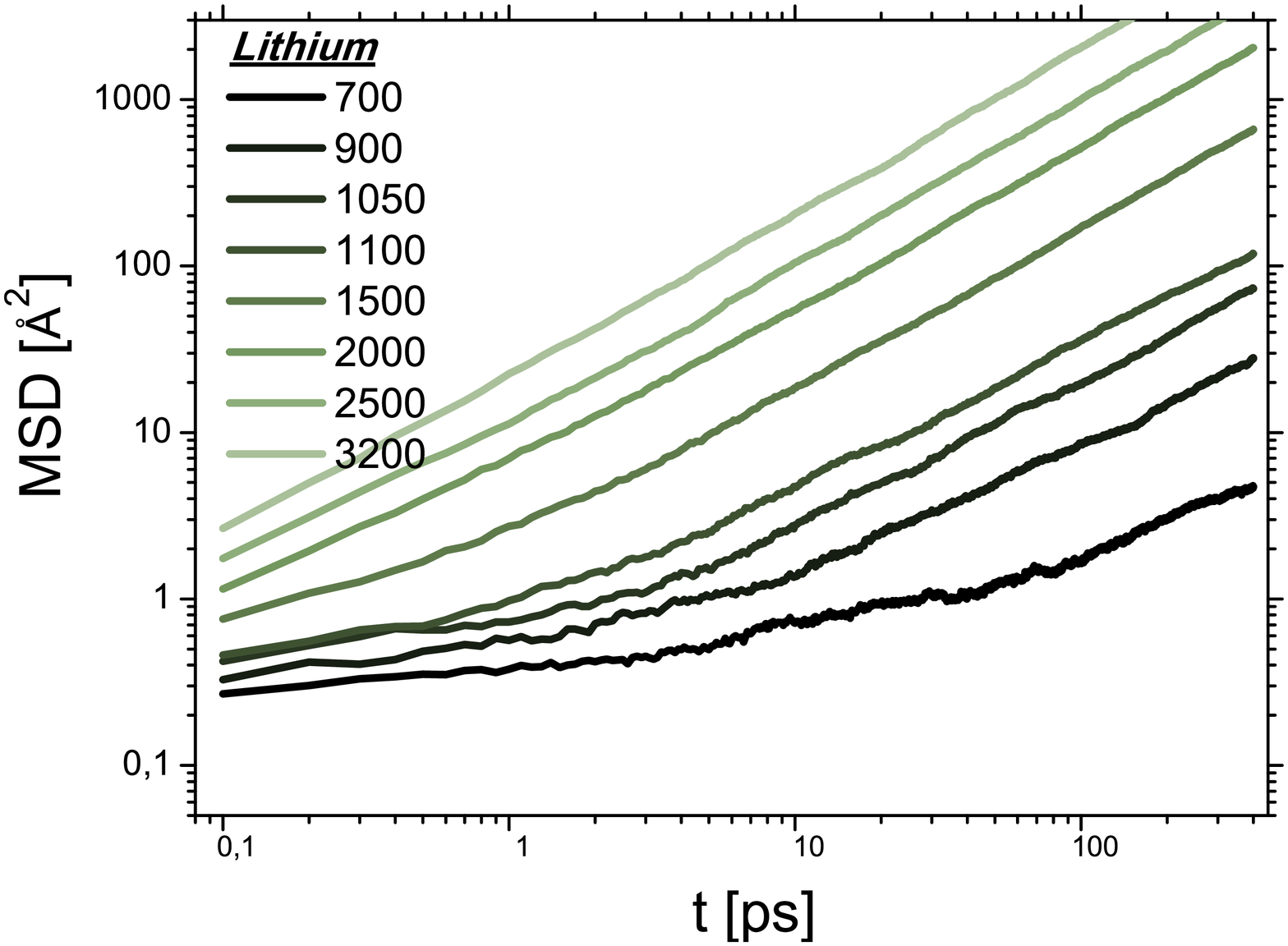}
\caption{Mean square displacement of $Si$, $O$ and $Li$ atoms  for the range of temperatures $600\,K\leq T\leq 3500\,K$. Silicon (left) and oxigen atoms 
(center) have a very similar behaviour, while lithium (right) diffuses much more than the other species at the same temperatures. }
\label{msd}
\end{figure*}

\begin{figure*}
\centering
\includegraphics[width = 0.85\columnwidth, height=0.6\columnwidth]{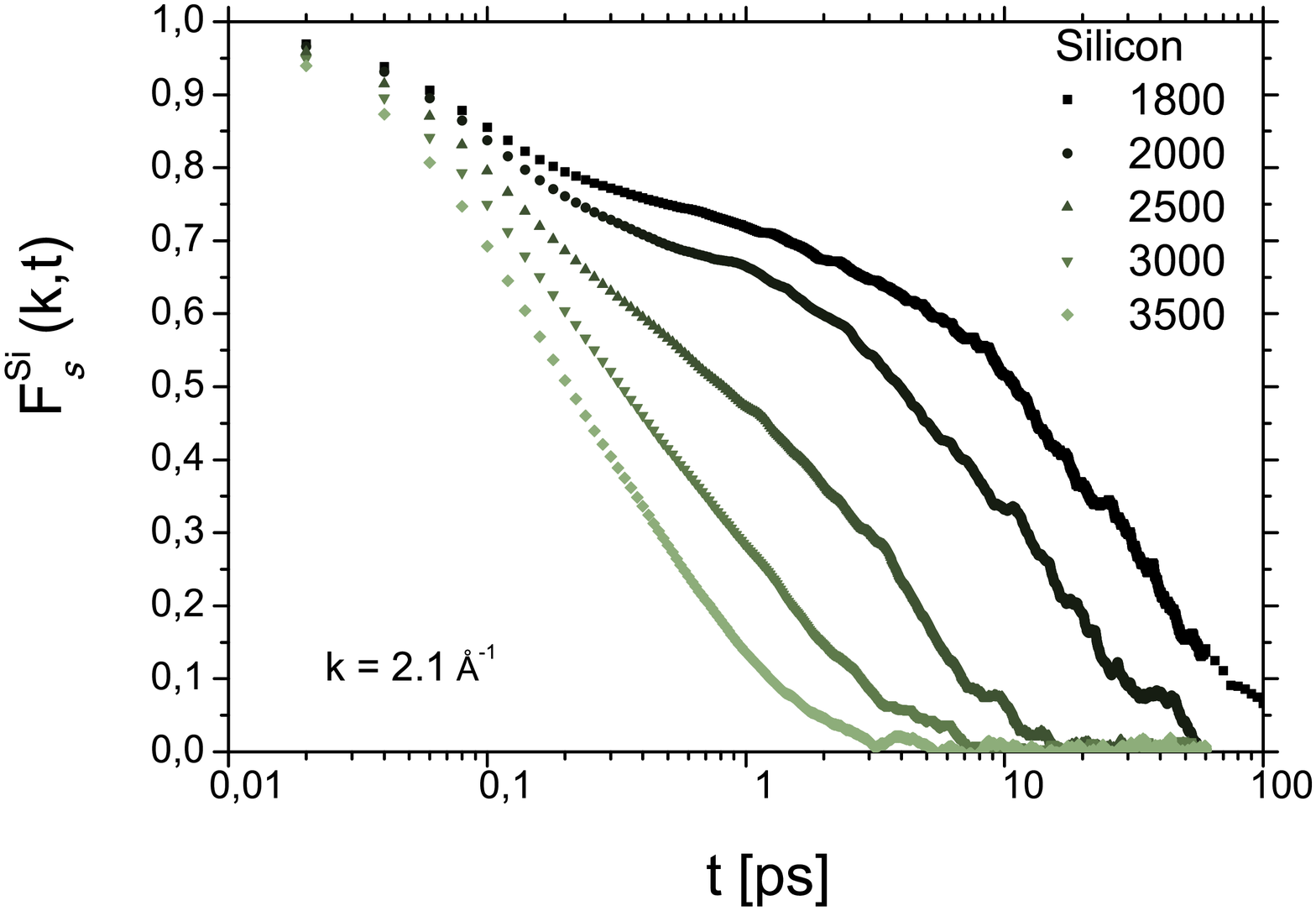}
\includegraphics[width = 0.85\columnwidth, height=0.6\columnwidth]{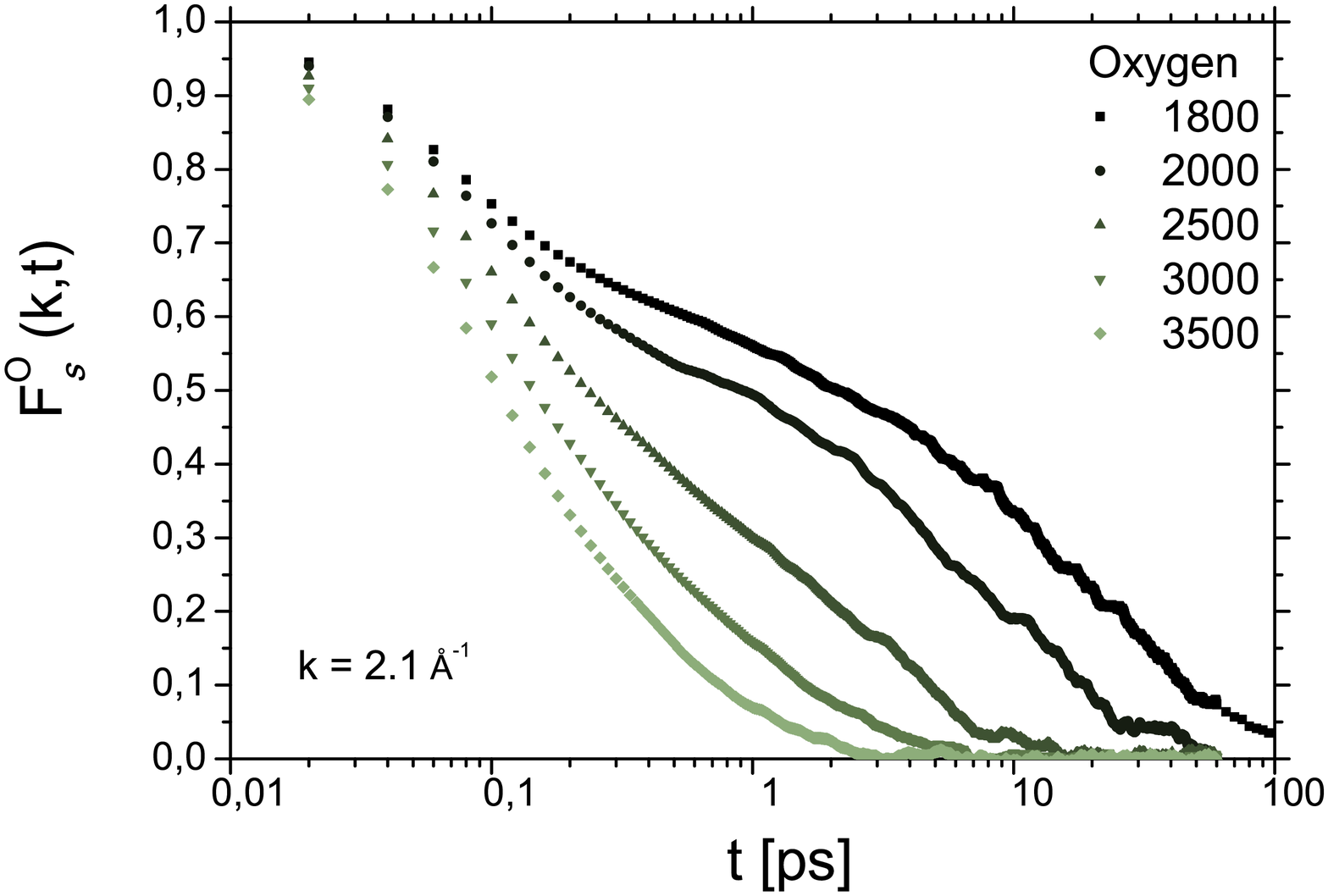}
\caption{Self intermediate scattering function of silicon (left) and oxigen (right) atoms for different temperatures and a common
wave vector (see text).}
\label{Fk}
\end{figure*}

\begin{figure}
\centering
\includegraphics[width = 1.0\columnwidth]{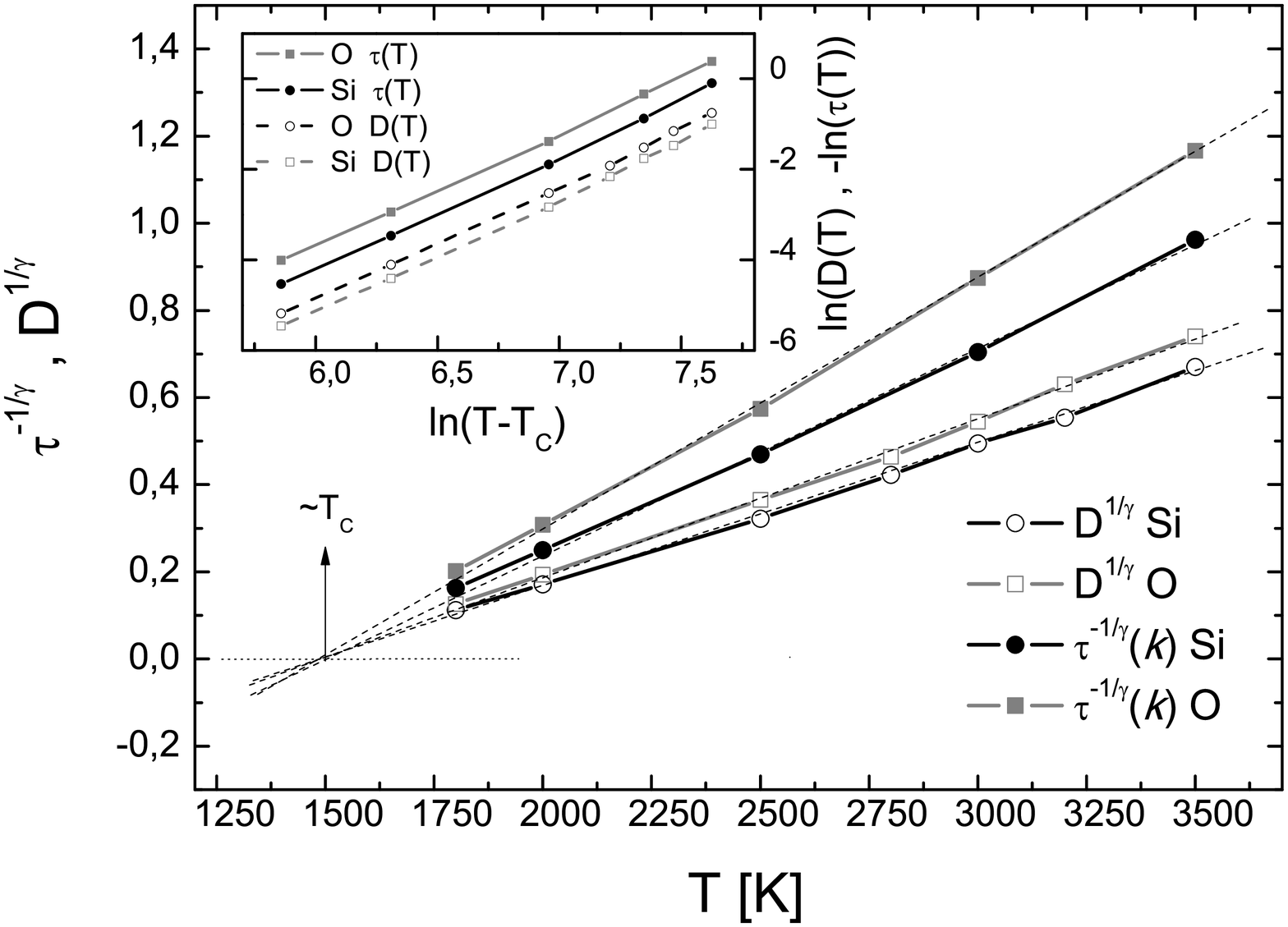}
\caption{Scaled diffusion constant and structural relaxation time of silicon and oxigen atoms versus temperature. The (extrapolated)
intersection defines the mode coupling critical temperature. Inset: the same data in a log-log scale which shows good agreement with
the power laws predictions of MCT (see text).}
\label{D_msd_vs_T}
\end{figure}

Changes in the structure of the system for the range of temperatures studied were monitored through
the radial distribution function $g(r)$. As can be seen in Fig. \ref{gr}(left), the RDF for Si-O has one large peak which corresponds to the distance 
between a silicon atom and its first neighbor oxigens that form the tetraedron  structure. Note that the Si-O RDF 
changes very little with temperature over the whole temperature range, signalling the robustness of the tetrahedra structure of the matrix. 
The RDF of 
Si-Si (Fig. \ref{gr}(center)) and Si-Li (Fig. \ref{gr}(right)) pairs have well defined peaks, reminiscent of a periodic solid structure.
From the behavior of the RDF one can conclude that there are not strong structural changes over the wide range of
temperatures studied, specially around the glass transition temperature $T_g$.
Inspite of this structural robustness the dynamical behaivor of the three types of atoms is very different. In these superionic glass-formers 
the alkali atoms move on a timescale which is, at low temperatures, many orders of magnitude faster than that for the atoms constituting the matrix 
(Si and O). In  Fig. \ref{msd} the mean squared displacement (MSD) of oxygen, silicon and lithium atoms is shown. 
Note that, on a time scale of $400\, ps$ lithium atoms reach diffusive behavior even at the lower temperatures, while the tetrahedra matrix ions 
remain frozen even at temperatures above $T_g$. 
It has been shown that several dynamical characteristics of ionic silicates can be explained within the framework of Mode Coupling Theory (MCT)
 \cite{HoKoBi2002,HoKo2002,HeKuVoBa2002}. When it applies, a distinctive prediction of MCT is a diffusional arrest at a characteristic temperature
$T_c$, which usually is above the corresponding glass transition temperature $T_g$. The theory predicts a power lw divergence of the 
structural relaxation time $\tau$ and a corresponding dropping of the diffusion constant to zero with the same power law, i.e. with a
common exponent $\gamma$:
\ba
D(T)& \propto & (T-T_c)^{\gamma} \\
\tau(T) & \propto & (T-T_c)^{-\gamma}
\ea
In order to extract the relaxation time we have computed the incoherent intermediate scattering function defined as:
\be
F_s^{\delta}(\vec k,t)=\frac{1}{N} \sum_{i=1}^{N_{\delta}} \langle \exp{[i\vec k \cdot (\vec r_i(t)-\vec r_i(0))]} \rangle
\ee
where the index $\delta$ labels the different species of atoms. For isotropic systems $F_s^{\delta}(k,t)$ depends only on the modulus
of the wave vector. In Fig. \ref{Fk}  the $F_s^{\delta}(k,t)$ for silicon and oxigen atoms are shown for different temperatures
in the high temperature regime. The wave vector was chosen to be $k=2.1\AA^{-1}$ 
which corresponds to the nearest neighbors distance between silicon atoms. The relaxation time $\tau$ was defined as
the time at which $F_s^{\delta}(k,t)$ decays to $0.1$. Fig. \ref{D_msd_vs_T}  shows the temperature dependence
of the difusion constant and the structural relaxation time for silicon and oxigen atoms. The inset of Fig. \ref{D_msd_vs_T} shows that $D(T)$ and
$\tau(T)$ obey a common power law scaling in this range of temperature, as  expected by MCT. 
Linear fits to these data give a value of the exponent $\gamma \approx 2.5$. We
verified that this value is robust for other values of the wave vector $k$. In the main figure we show that extrapolation of the data to
low temperatures results in an intersection point of all data sets at a characteristic temperature $T_c \sim 1500\,K$, which corresponds to
a dynamic arrest of the Si and O atoms. Then, we associate this $T_c$ with the mode coupling critical temperature of the system, as discussed above.
We conclude this section noting that it is possible to identify at least two characteristic temperatures of the lithium metasilicate 
matrix where the dynamics suffer qualitative changes: the glass transition temperature $T_g \simeq 1000\,K$ and the (higher) mode coupling critical temperature
$T_c \sim 1500\,K$. We have also seen that no structural signal of these dynamical transition temperatures is observed in the radial correlation
functions. In the next section we seek for possible structural signatures of these transitions by looking at two different orientational order parameters.

\section{Orientational measures} \label{sec.orientational}
\subsection{Local orientational order around $Si$ atoms}
Given that characteristic dynamical temperatures does not seem to be correlated with structure as given by the radial distribution function,
we performed a search for this correlation by looking at some orientational measures. 
Local orientational order can be determined by the Steinhardt or bond orientational order parameters (BOO) defined as
\cite{Steinhardt1983}:
\be \label{Qparameter}
Q_l = \left[\frac{4\pi}{2l+1}\sum_{m=-l}^{l} |\bar{Q}_{lm}|^2\right]^{1/2},
\ee 
where $\bar{Q}_{lm} = \langle Q_{lm}(\vec r) \rangle$ and 
$Q_{lm}(\vec r)= Y_{lm}(\theta(\vec r),\phi(\vec r))$ are spherical harmonics with $\theta(\vec r)$, $\phi(\vec r)$ the polar angles of a bond measured with respect to some reference coordinate system and where the average is taken over some suitable set of bonds. 
A ``bond'' is a line that connects the centers of two neighboring atoms. The matrix of tetrahedra in $Li_2SiO_3$ is disordered in the sense of 
simple crystalline order, and we do not expect to obtain meaningful information computing BOO parameters beyond first or second neighbors. 
Nevertheless, they can give valuable information on the evolution with temperature of short range order, i.e. tetrahedra. 

\begin{figure}
\includegraphics[width = 1.0\columnwidth]{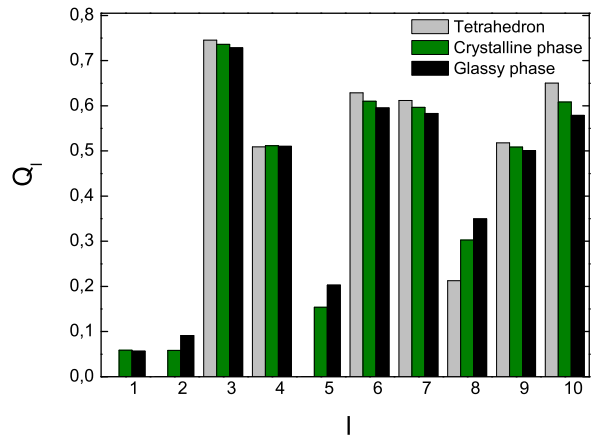}
\caption{Bond orientational order parameters $Q_i$ for $i=1,\ldots, 10$ comparing a perfect tetrahedron, a tetrahedron from
the crystalline cell of the system and one in the low temperature glass phase  at $T=600\,K$.}
\label{Qis}
\end{figure}

\begin{figure*}
\centering
\includegraphics[width = 0.9\columnwidth, height=0.6\columnwidth]{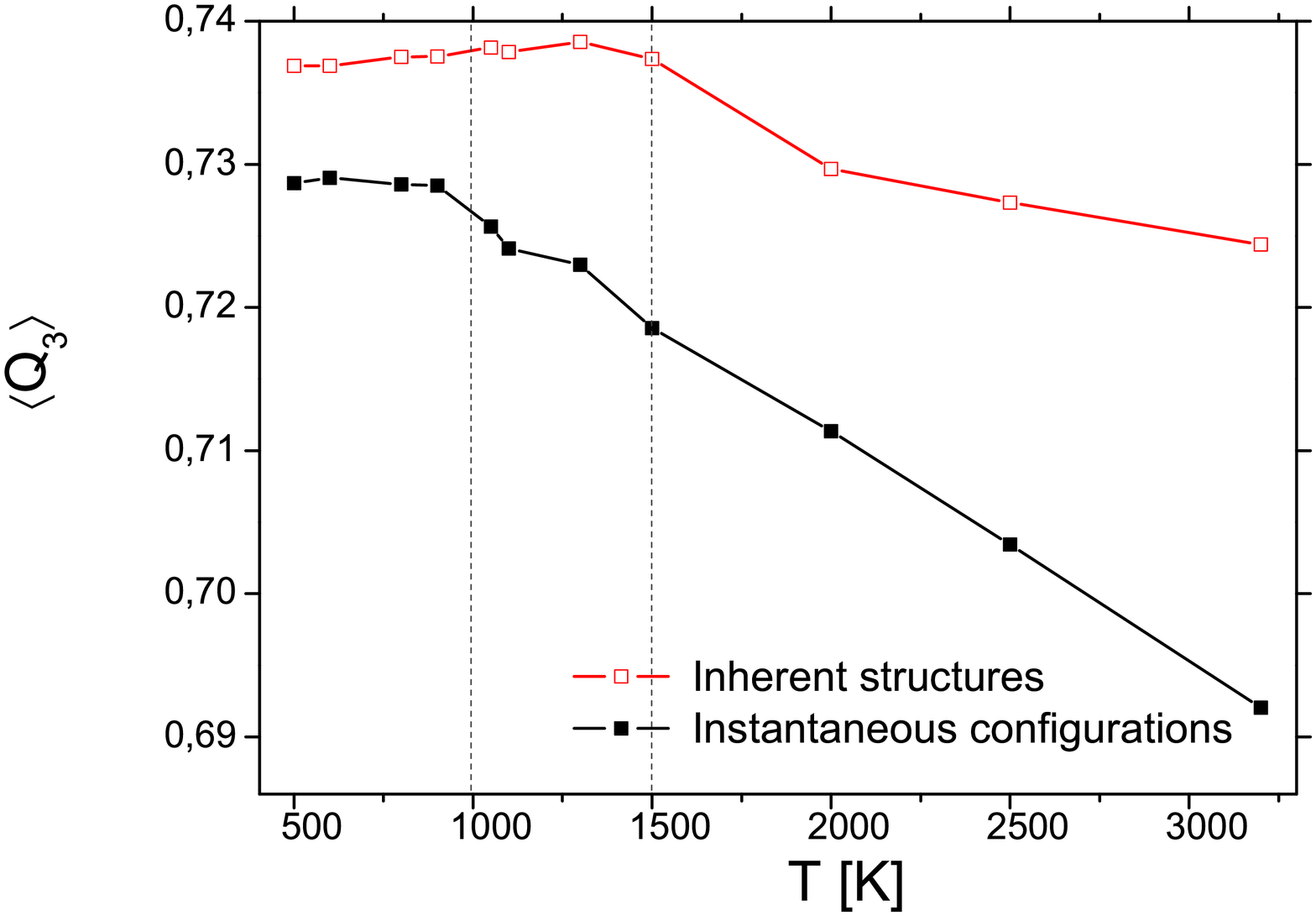}
\includegraphics[width = 0.9\columnwidth, height=0.6\columnwidth]{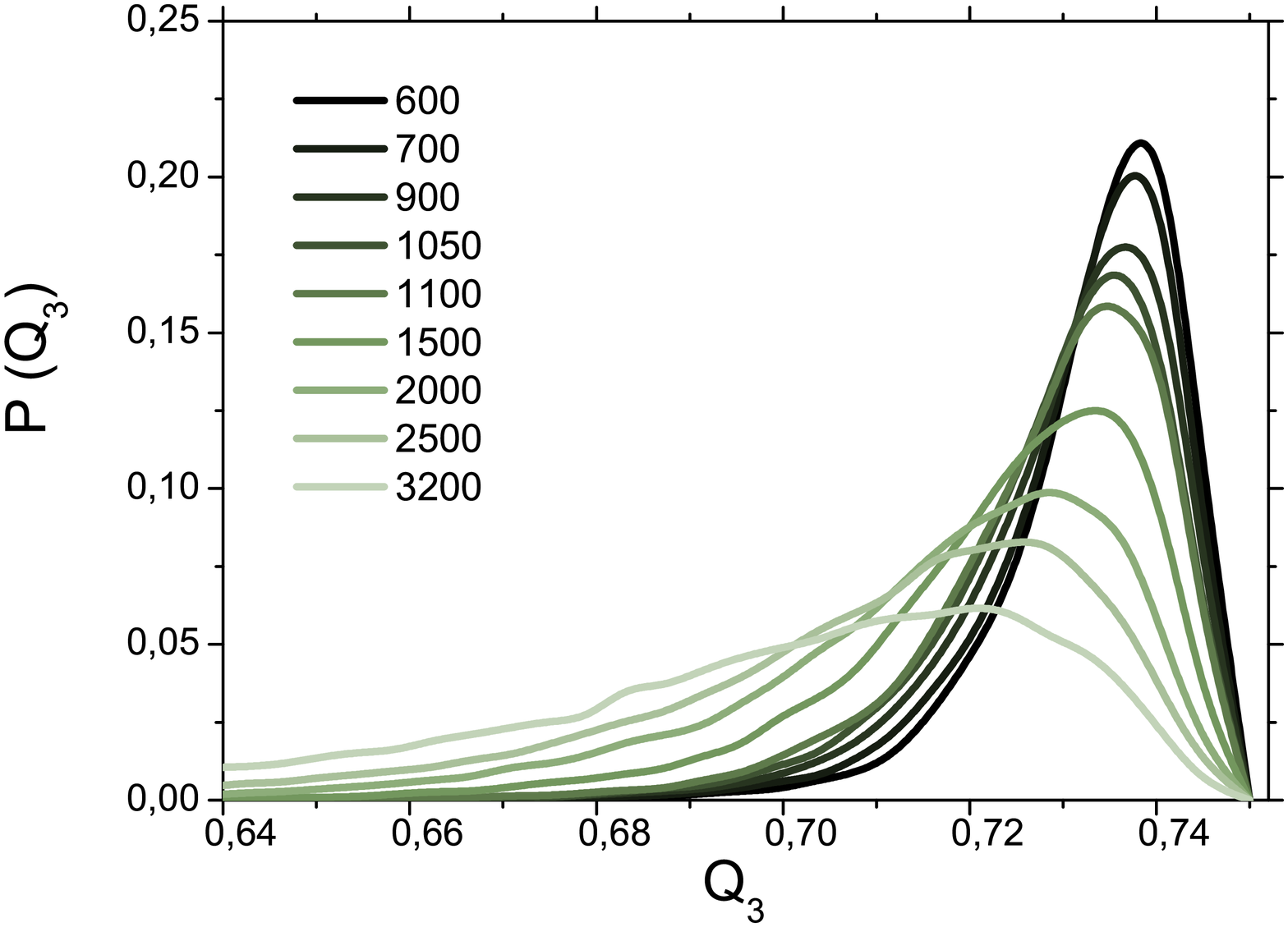}
\caption{Left: Temperature dependence of $Q_3$ for instantaneous (black) and inherent structure (red) configurations.
Right: distribution of instantaneous $Q_3$ values for different temperatures.}
\label{dist_Q3}
\end{figure*}

In this work we considered BOO of the neighborhood of $Si$ atoms, as determined by the first peak of the $g(r)$ of $Si$-$O$.
Fig. \ref{Qis} shows a bar plot with the values of $Q_l$ from $l=1$ to $l=10$ comparing three different systems: a perfect tetrahedron, the crystalline structure of the system \cite{TaLu2010} and the glass structure of metasilicate at $T=600K$. 
The small differences in $Q_l$ values between a perfect tetrahedron and $SiO_4$ tetrahedra in silicates is because in the
last the tetrahedra are not mutually independent (isolated), most are linked by bond oxygens which distort somewhat the optimal structure.
The existence of such BO type impedes the perfect orientation of individual tetrahedra, even in the crystalline phase.
The dominant symmetry is given by $l=3$ which reflects the 3-fold symmetry of thetrahedra~\cite{WaSt1991,SaAn2003}. 
In Fig \ref{dist_Q3} we show how $\langle Q_3 \rangle $ evolves with temperature. 
Averages were taken from 30 different configurations of the whole system to improve statistics. In each configuration 576 tetrahedra (number of silicons) were analyzed, resulting in a total of 17280 tetrahedra at each temperature.
In Fig \ref{dist_Q3}(a) the lower (black) points correspond to instantaneous configurations along
a molecular dynamics trajectory and the upper (red) points correspond to measures in the inherent structures from equilibrium configurations
at each temperature. It can be seen that, up to approximately $T \sim 900\,K$ both data sets behave similarly and show no variation with
temperature. The system is effectly frozen in its glass phase in this regime. Above $T \sim 900\,K$ orientational order of instantaneous
configurations decreases, signalling a change in the tetrahedron structure around $Si$ atoms. Nevertheless, the corresponding inherent
structures measures show a constant value up to $T \sim 1500\,K$. This means that tetrahedra in the interval $900\,K\leq T \leq 1500\,K$
are robust structures, i.e. they can deform and accomodate in order to find more stable global configurations of the network, but the 
landscape is characterized by a common tetrahedral strucutre in all this temperature range. In Fig \ref{dist_Q3}(b) the 
distribution of instantaneous $Q_3$ values is shown for a wide range of temperatures. 
The distribution is peaked around the mean value at low temperatres and broadens,
as expected, as temperature grows, but otherwise we can see no obvious signs of structural changes from the form of the distribution in
a wide temperature range. This spread in the distribution of $Q_3$ was  previously observed for silicon systems in its supercooled phase 
\cite{SaAn2003} and used to quantify the modification of the local structure when the system experiences a phase transition.
Note that, for temperatures as high as $2500\,K$ the probability to find a tetrahedral structure of oxigens
around a silicon atom is relatively high, i.e. tetrathedra are very robust structures.
\subsection{Relative orientations between tetrahedra}
\begin{figure}
\includegraphics[width = 1.0\columnwidth]{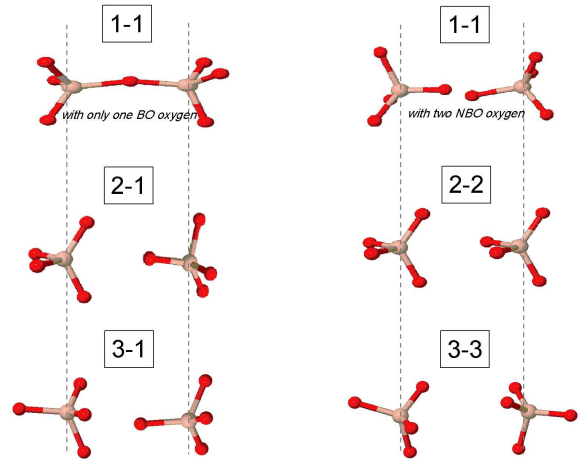}
\caption{Defintion of the {\it classes} of the Rey's parameters. Possible relative orientations of a couple of tetrahedra showing the number of shared oxigens of the two silicons.
Note that there are two possibilities in which two silicons share one oxigen each: a common BO and two NBO's (see text).}
\label{ReyEx}
\end{figure}

Note that the BOO parameter $Q_3$ described in the previous section captures the local orientational order between a $Si$ atom and the $O$
atoms sorrounding it. In the system under study, this amounts to characterize the geometry of single tetrahedra for different temperatures.
If one wants to go further in the characterization of orientational order it is important to understand how the tetrahedra correlate in space,
 how they organize relative to each other when the temperature changes. 
In general, if local order is simple, as is the case in simple molecular liquids forming highly symmetric crystals at low
temperatures, then the spatial correlations of BOO parameters can be used to search for medium and long range order in the system. 
In the case of $Li_2SiO_3$, which forms a complex silicate network of tetrahedra modified by lithium ions, a useful  correlation function
 of the local order parameter should capture the {\it relative orientation between tetrahedra}. To quantify this relative orientational order in 
tetrahedra forming systems Rey has introduced a useful quantity for the case of carbon tetrachloride  \cite{Rey2007} which we here adapt to 
ionic silicates. 
 The Ray parameters (RP) are defined as follows: take two silicon atoms $Si_1$ and $Si_2$ and define the vector $\vec r_{12}$ that connects them. 
Then define the two planes passing through the $Si$ atoms and perpendicular to $\vec r_{12}$. Then count how many oxigen atoms are shared by
the two $Si$ atoms between these 
planes and define different ``orientational classes''. Note that, due the tetrahedral character of the network, there are a few possibilities
of oxigen sharing configurations: if there are two oxigens between the planes, this is the 1:1 ``corner-to-corner'' configurations. If there
are three $O$'s then this is the 2:1 or ``corner-to-edge'' configurations, and so on. At most, the two $Si$ atoms can share three oxigens
each, corresponding to a face of the tetrahedron. This is the 3:3 or ``face-to-face'' configuration.
 In the Fig \ref{ReyEx} we show some possibilities of configurations and their respective names. 
Each coniguration defines one class of the Rey's parameters.
Different from carbon tetrachloride in which
each chlorine atom is connected with a single carbon, in $Li_2SiO_3$ a fraction of oxigens are ``bonding'', i.e. are at a vertex connecting two
tetrahedra. Then, we adapted the original definition in order to acommodate this possibility: Bonding oxigens are considered to belong to 
both $Si$ atoms, and then counted twice in the corresponding cases.
\begin{figure}
\includegraphics[width = 1.0\columnwidth]{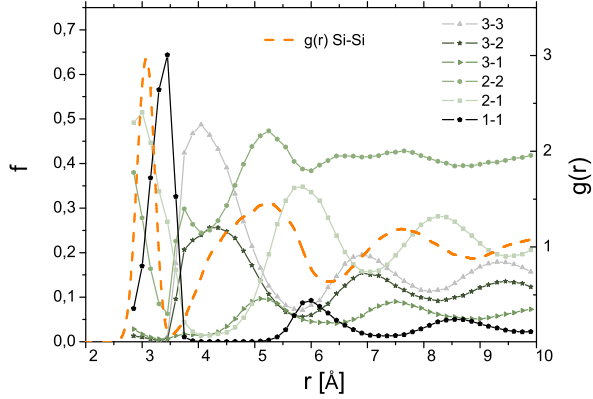} 
\caption{Percentages of each class of Rey's parameters as function of distance at $T=1500\,K$. The RDF of Si-Si is shown (dashed) to give 
an idea of the distances between the silicon atoms that are at the center of the tetrahedra. }
\label{Rey1}
\end{figure}
Figure \ref{Rey1} shows the percentage of each class, $f$, as a function of the distance between $Si$ atoms for  $T=1500\, K$, together
with the radial distribution function $Si$-$Si$. Within the first peak of the RDF there appear only three classes with a finite contribution:
corner-to-corner (1:1), corner-to-edge (1:2) and edge-to-edge (2:2). Up to the second maximum of the RDF the relative percentages of the
different classes fluctuate considerably. Nevertheless, at short distances configurations with a few sharing oxigens are favored. 

\begin{figure}
\includegraphics[width = 1.0\columnwidth]{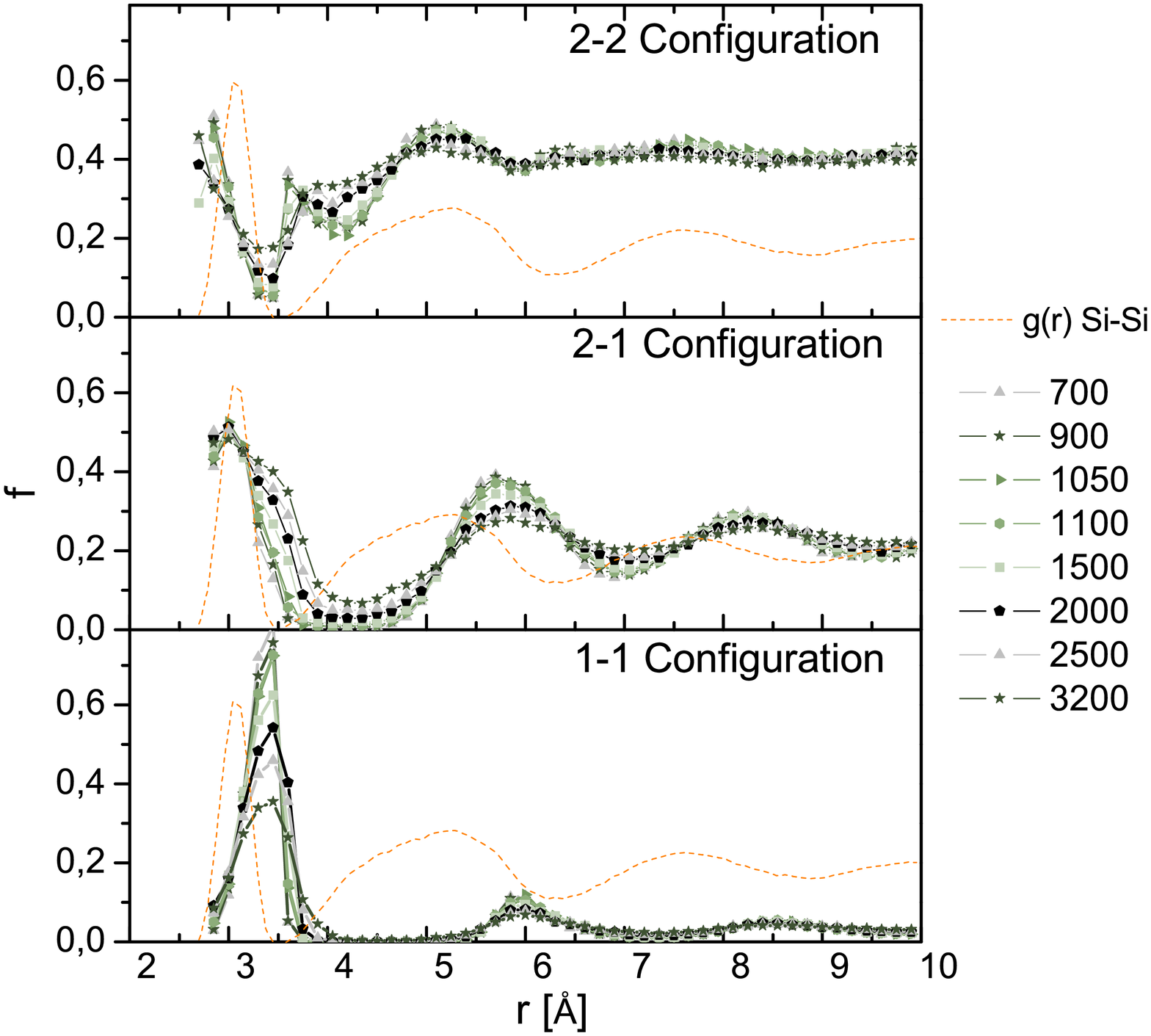} 
\caption{The most frequent Rey's configurations near the first peak of the RDF for different temperatures.}
\label{ReyT}
\end{figure}

\begin{figure}
\includegraphics[width = 1.0\columnwidth]{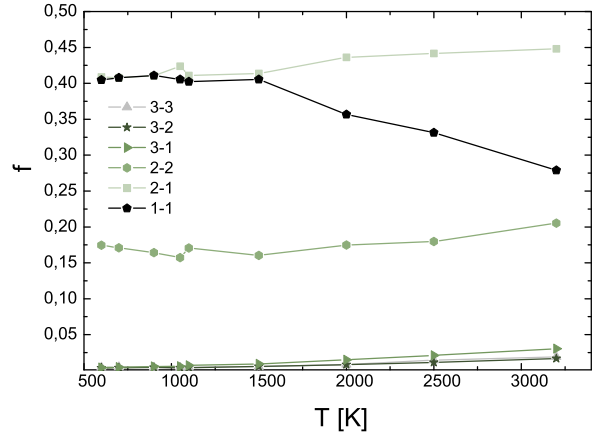}
\caption{Variation with temperature of the percentages $f$ of the Rey's parameters for pairs of silicon atoms at a distance corresponding to the first peak of the 
$Si-Si$ RDF (nearest-neighbors atoms).}
\label{short-range-1}
\end{figure}

\begin{figure}
\includegraphics[width =1.0\columnwidth]{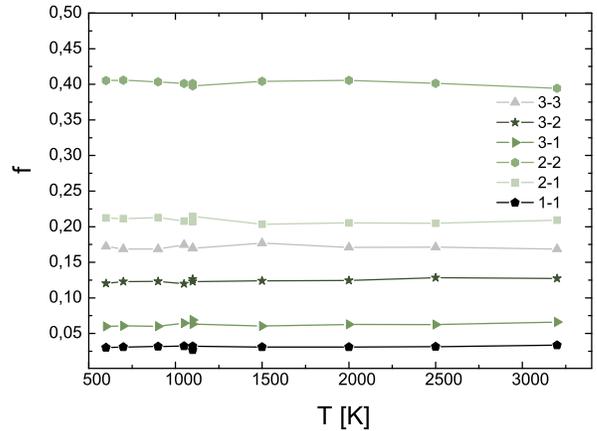}
\caption{Variation with temperature of the percentages $f$ Rey's parameters corresponding to the second peak of the $Si-Si$ RDF.}
\label{short-range-2}
\end{figure}

In Figure \ref{ReyT} the three dominant configurations near the first peak of the RDF are shown as function of distance between silicon
atoms for different temperatures. It can be seen that there are not
strong qualitative variations with temperature except for a change in the height of the peaks.
Next, we address the behavior of RP for $Si$ atoms at the first peak of the RDF (nearest-neigbors) and also for 
the second (broad) maximum (second neighbors). Figure \ref{short-range-1} shows the temperature dependence of all
classes for two nearest-neighbors $Si$ atoms. It is evident that for two nearest-neighbors tetrahedra at any temperature 
$600\,K\leq T\leq 3200\,K$ the only configurations with sizeable probability are 1:1, 2:1 and 2:2. Furthermore, 
classes 1:1 and 2:1 are more probable than 2:2 at any temperature. At $T \approx 1500\,K$ an interesting bifurcation
in the probabilities of the two more probable classes occurs. It is seen that coming from high temperatures where the
probabilities of each class approach those of a random configuration of tetrahedra, the three more probable configurations
change continuously until $T \sim 1500\,K$, below which no more changes in the values of any of the classes is observed.
This points to a freezing of the RP's below $T_c \sim 1500\,K$, which corresponds with the mode coupling critical temperature of the
system. Below this temperature any two nearby tetrahedra keep their relative orientations frozen, no more 
orientational rearranges occur in the network. Note that, although this points to a freezing of local orientational structure
at $T_c \sim 1500\,K$, the BOO parameter $Q_3$ continues growing until a much lower temperature $T\sim 900\,K$, near the glass
transition temperature.
In Figure \ref{short-range-2} the evolution with temperature of RP's is shown for two $Si$ atoms at a distance
corresponding roughly to the second maximum of the RDF. The difference with the nearest-neighbors case is striking. One can 
note that in this case all classes show an approximately constant value in the whole temperature range. The values correspond 
roughly to a random distribution, or a ``gas'' of tetrahedra. For the $Li_2SiO_3$ system tetrahedra are never completely
random because of the presence of bonding oxigens. 
\section{Discussion and Conclusion} \label{sec.conclusions}
\begin{figure}
\includegraphics[width =1.0\columnwidth]{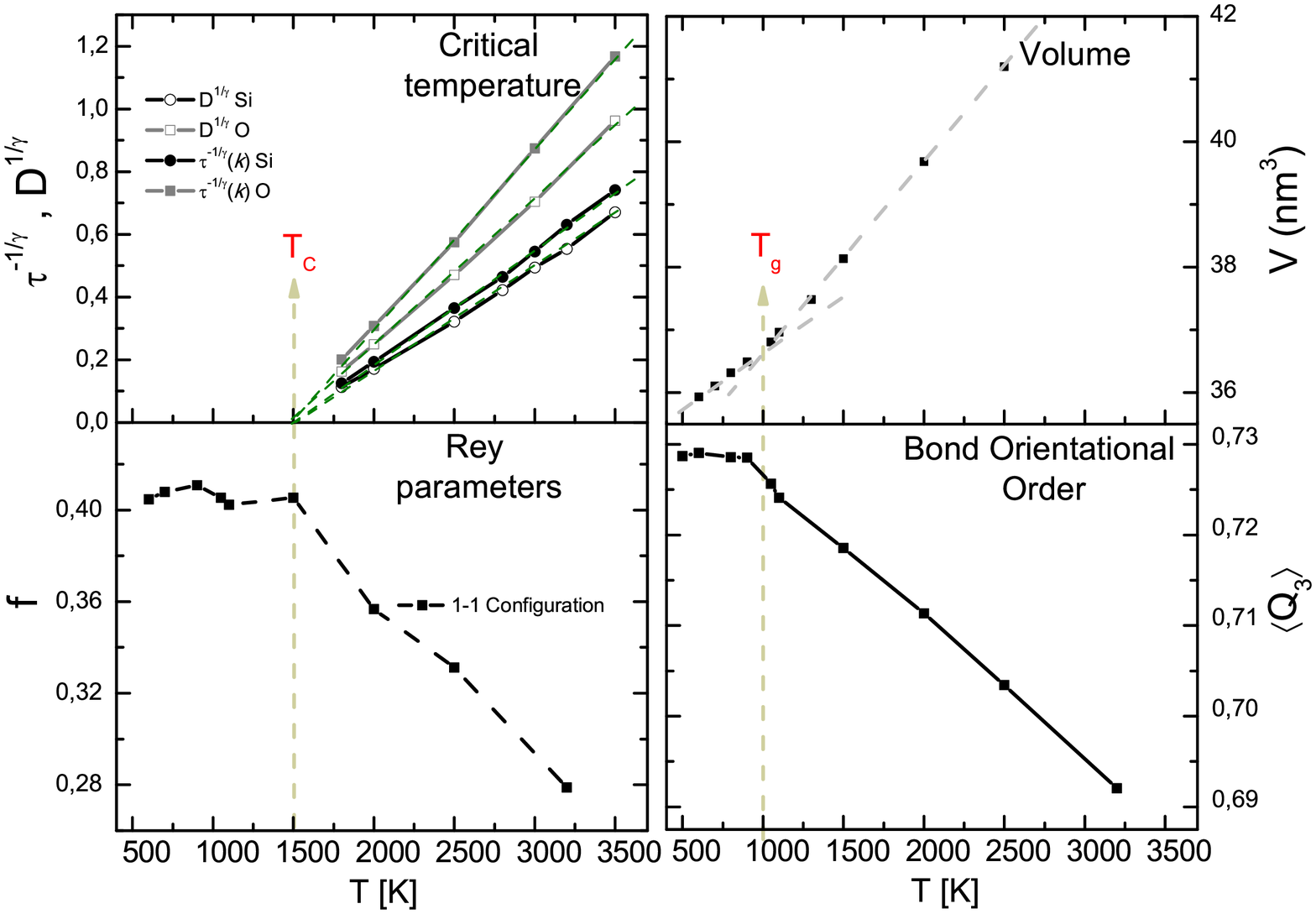}
\caption{The four plots illustrate the presence of the two characteristic temperatures $T_c$ and $T_g$ at which dynamical and
thermodynamical quantities show some sort of singularities, as explained in the text.}
\label{comparative}
\end{figure}

The main result of our study is the identification of two characteristic temperatures where both dynamical and
structural changes in the lithium metasilicate matrix are evident. The results described in the previous sections are
summarized in Figure \ref{comparative}, where we show four plots comparing the outcome of dynamical and thermodynamical
measures. 
First of all, from mean squared displacement and dynamical self-correlation functions we
were able to determine a characterstic temperature Tc, the critical mode coupling temperature.
We find that these quantities are well described by  the mode coupling  predictions,
particularly  the existence of a power law behavior of diffusivity and
relaxation times with temperature with a common exponent near the (theoretical) critical one $T_c$.
This is shown in Fig \ref{comparative} top-left panel.
The top-right panel of the figure shows the usual characterization of the glass transition temperature (for the same network forming species)
from the volume versus temperature plot. 
The analysis of the orientational measures as a function of temperature (in the bottom panels of the Fig \ref{comparative}) allow us
 to understand what happens microscopically in the system. Below $T_c$, relative movement between tetrahedra ceases, although the system keeps 
relaxing until $T_g$. This means that bellow $T_c$ the main relaxation processes are not related to rotation of tetrahedra; they cannot 
be seen as rigid bodies that perform rigid rotations. Instead, the atoms that form the tetrahedra keep relaxing until $T_g$.

Worth noting is the large size of the temperature window between the mode
coupling critical temperature and the glass transition one. Many studies on the mode coupling phenomenology associate
$T_c$ with a transition at which the lanscape probed by the system suffers a topological change \cite{AnDiRuScSc2000,
GrCaGiPa2002} from a regime in which the dynamics is strongly influenced by the proximity of saddle points in configuration
space to a regime below $T_c$ where the dynamics is governed by activated processes over energy barriers.
Although the existence of the high temperature regime has been characterized in many cases and found to be in qualitative
agreement with the predicitions of MCT, the low temperature regime has been much less studied, mainly due to the strong
dynamic arrest experienced by the systems below $T_c$. At this point it is important to stress that MCT has been mainly
tested in simple glass formers, like binary mixtures, which do not show short range structures like network-forming
systems and where $T_c$ and $T_g$ are hardly distinguishable. We have shown that lithium metasilicate shows both 
transitions to be separated by a large temperature window. This suggests that these kind of glass formers may be
interesting systems to study details of the dynamics beyond MCT, specially the nature of the elusive ``activated events''.
Also much discussed today is the old question about thermodynamic signatures of the glass transition, like growing
static correlations \cite{BerthierRMP2011, BiroliBouchaud2012}. While it is clear that there are no ``simple'' structures
associated with the dynamical arrest near $T_g$, the role played by orientational order and correlations has been much
less studied in this context \cite{tanaka2010,Tanaka2012}. Here we have shown that, indeed, there seems to be a clear
correlation between dynamical arrest and orientational order. This is summarized in the two bottom panels of Figure
\ref{comparative}. Two order parameters which measure the degree of local tetrahedral structure and the relative orientation
between neighbor tetrahedra show a distinctive change in behavior with temperature, at approximately the same temperatures
characterizing the mode coupling and glass transitions, $T_c$ and $T_g$, respectively. This is the main result of this
study and it suggests that it is possible to characterize the glass transitions through sharp changes in orientational parameters.
While the case of alcali silicates is probably an extreme one, in the sense that the network of tetrahedra is extremely
disordered due to the effect of alcali atoms on the network, the present approach can probably lead to
interesting insights on the interplay between dynamics and order in other families of network forming systems which show different
degrees of ordering, e.g. plastic glasses \cite{NCaballero2012,Angell2013}. By monitoring the type and
relative abundance of selected atoms at the vertices of tetrahedra the connection between dynamics and structure can 
be studied in detail and may allow to gain important insights in the nature of glassy behavior of complex systems. 
\acknowledgments
The {\em Conselho Nacional de Desenvolvimento Científico e Tecnológico} (CNPq, Brazil) and the {\em Consejo de Investigaciones Científicas y Técnicas} (CONICET, Argentina) are aknowledged for partial financial support. 

\end{document}